\documentclass{article} %
\usepackage{iclr2022_conference,times}

\usepackage{amsmath,amsfonts,bm}

\def\eqref#1{equation~\ref{#1}}

\def\1{\bm{1}}

\DeclareMathAlphabet{\mathsfit}{\encodingdefault}{\sfdefault}{m}{sl}
\SetMathAlphabet{\mathsfit}{bold}{\encodingdefault}{\sfdefault}{bx}{n}

\usepackage{hyperref}
\usepackage{url}
\usepackage{natbib}
\usepackage{amsmath,amssymb,amsfonts}
\usepackage{mathtools}
\usepackage{cleveref}
\usepackage{booktabs}
\usepackage{soul}
\usepackage{color}
\usepackage{dsfont}
\usepackage{balance}
\usepackage{siunitx}
\usepackage{subcaption}
\usepackage{tikz}

\definecolor{ForestGreen}{RGB}{21, 155, 82}
\definecolor{Dandelion}{RGB}{249, 180, 43}

\DeclareRobustCommand\sampleline[1]{%
  \tikz\draw[#1] (0,0) (0,\the\dimexpr\fontdimen22\textfont2\relax)
  -- (1em,\the\dimexpr\fontdimen22\textfont2\relax);%
}
\renewcommand{\headrulewidth}{0pt}
\title{NeuRL: Closed-form Inverse Reinforcement Learning for Neural Decoding}

\author{Gabriel Kalweit$^{1,2,}$\thanks{Corresponding author (\texttt{kalweitg@cs.uni-freiburg.de}).}\phantom{ }, Maria Kalweit$^{1,2}$, Mansour Alyahyay$^{2,3}$,\\\textbf{Zoe Jaeckel$^{2,3}$, Florian Steenbergen$^{2,3}$, Stefanie Hardung$^{2,3}$,}\\\textbf{Thomas Brox$^{2,4}$, Ilka Diester$^{2,3,5}$ and Joschka Boedecker$^{1,2}$}\\
$^1$Neurorobotics Lab, Department of Computer Science, University of Freiburg, Germany\\
$^2$Cluster of Excellence BrainLinks-BrainTools//IMBIT, University of Freiburg, Germany\\
$^3$Optophysiology Lab, Department of Biology, University of Freiburg, Germany\\
$^4$Computer Vision Group, Department of Computer Science, University of Freiburg, Germany\\
$^5$Bernstein Center Freiburg, University of Freiburg, Germany}

\begin{document}

\maketitle

\begin{abstract}
Current neural decoding methods typically aim at explaining behavior based on neural activity via supervised learning. However, since generally there is a strong connection between learning of subjects and their expectations on long-term rewards, we propose NeuRL, an inverse reinforcement learning approach that (1)~extracts an intrinsic reward function from collected trajectories of a subject in closed form, (2) maps neural signals to this intrinsic reward to account for long-term dependencies in the behavior and (3) predicts the simulated behavior for unseen neural signals by extracting Q-values and the corresponding Boltzmann policy based on the intrinsic reward values for these unseen neural signals. We show that NeuRL leads to better generalization and improved decoding performance compared to supervised approaches.  We study the behavior of rats in a response-preparation task and evaluate the performance of NeuRL within simulated inhibition and per-trial behavior prediction. By assigning clear functional roles to defined neuronal populations our approach offers a new interpretation tool for complex neuronal data with testable predictions. In per-trial behavior prediction, our approach furthermore improves accuracy by up to 15\% compared to traditional methods.
\end{abstract}

\section{Introduction}
\label{sec:intro}

Neural decoding methods use neural spiking activity from the brain to infer predictions about behavior, like explaining or predicting movements based on activity in the motor cortex \citep{Peixoto2021,Melbaum2021.03.04.433869,ba9e210f50474bfd834a22daae0898f0} %
or decisions based on activity located in prefrontal and parietal cortices \citep{BAEG2003177,ibos2017sequential}. Decoding can be used to control brain machine interfaces \citep{interface2, interface1, interface3} or to extract general working principles of the brain. %
Recently, deep learning has shown great potential in a number of domains %
and is outperforming classical approaches %
in the field of neural decoding  \citep{neuraldec,GLASER2019126}. Nevertheless, decoding methods are usually trained supervised for prediction \citep{10.3389/fncir.2019.00075,8852121}, mapping greedily from neural signals directly to actions without reasoning about the long-term consequences of the actions. In the reinforcement learning (RL) paradigm, on the other hand, this is accounted for explicitly by learning a policy which maximizes long-term rewards in expectation. Prior work also showed that learning in the brain is driven by changes in the expectations about rewards and punishments \citep{Schultz1997ANS} which naturally aligns with the RL framework. Importantly, the immediate reward function in RL can be seen as the most succinct, robust, and transferable definition of behavior to be learned \citep{Abbeel:2004:ALV:1015330.1015430}. 
Consequently, in this work, we propose the use of \emph{inverse reinforcement learning} (IRL) methods to infer an intrinsic reward function explaining  observed animal behavior, allowing us to draw conclusions about neural activity and its relation to the recorded behavior, as well as improving generalization and decoding performance.

\begin{figure}[t]
    \centering
        \includegraphics[width=\textwidth]{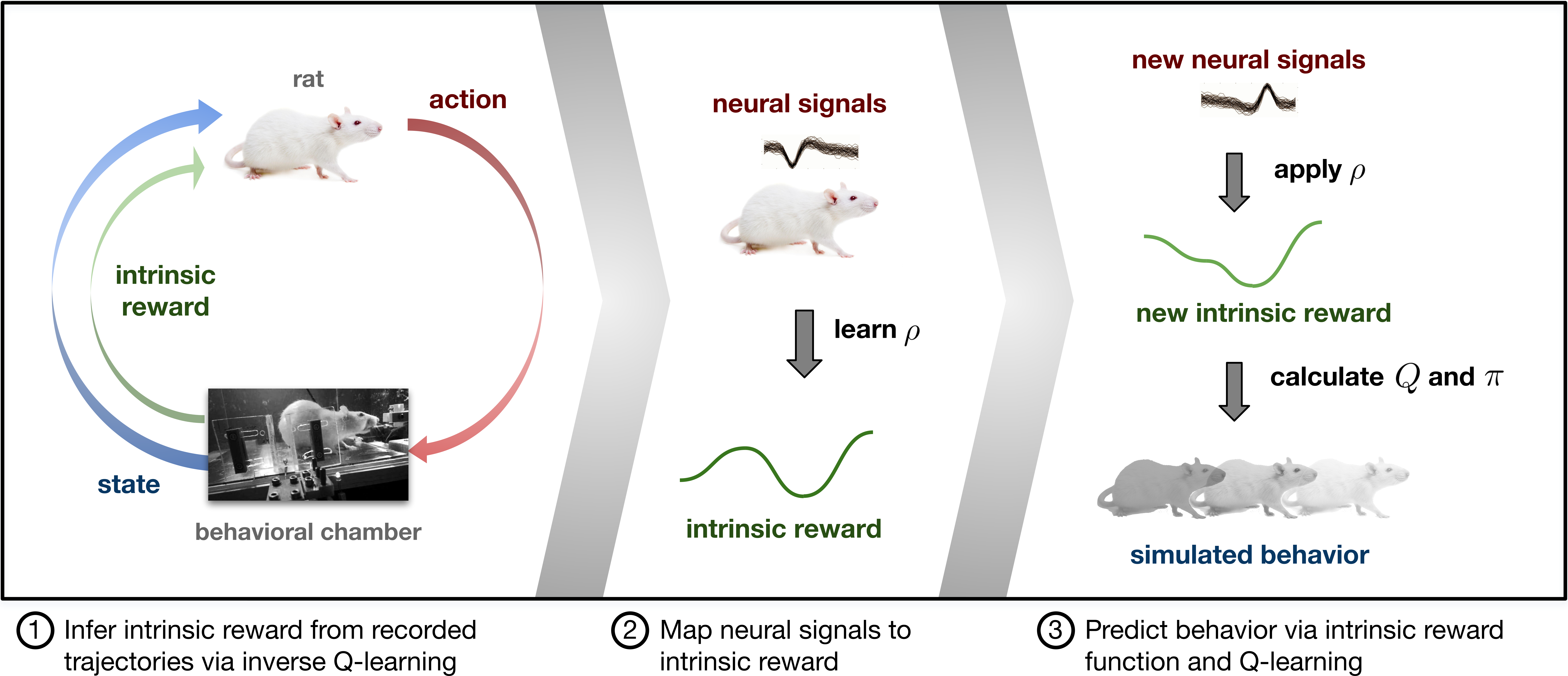}
    \caption{Response-preparation task in a reinforcement learning setting. A rat acts in a behavioral chamber with a lever and a sugar port. Our proposed framework first infers an intrinsic scalar reward function of the rat's behavior via closed-form inverse reinforcement learning. Then, a parameterized function $\rho$  is learned which maps neural signals to the intrinsic reward. Finally, we generalize to new situations by applying $\rho$ to other neural signals and calculating the Q-values and Boltzmann policy to study the corresponding simulated behavior for these neural signals.}
    \label{fig:ratrlsetting}
\end{figure}

We use Inverse Action-value Iteration (IAVI) \citep{NEURIPS2020_a4c42bfd} to calculate the immediate reward function analytically in closed-form assuming that a demonstrator is following a Boltzmann distribution over its unknown optimal action-values which in turn represent the expected long-term return for observed actions. This common assumption has already been applied to model the behavior of humans and animals in a plethora of prior work \citep{bitterman1965phyletic,Baker2007,SilvaNature}. The learned reward function formalized in IAVI encodes the local probabilities of the demonstrated actions while enforcing the local probabilities of the maximizing actions in the future under Q-learning. In contrast, common supervised learning methods only consider the action taken in the current time step. In this work, we propose to instead estimate a mapping of recorded neural signals to the immediate reward function learned on observed rat trajectories via IRL as an intermediate step to find coherences between neural spikings and taken actions.
The learned mapping can then be used to calculate the intrinsic reward for unseen neural signals and simulate a rat's behavior based on the new reward, an approach we call \emph{NeuRL}. The scheme of the algorithm is shown in \Cref{fig:ratrlsetting}. This decoding mechanism can be used to predict behavior in real-time from neural spiking or it can simulate the influence of specific neurons on the behavior of the rat. Our proposed decoding tool can help to interpret complex neural data and can serve as a hypothesis generator which then can be evaluated in vivo.

We study the behavior of rats in a response-preparation task where rats ought to hold a lever until a cue (vibration to the paw) indicates that the animal should release. \Cref{fig:ratsinchamber} shows a rat performing this task in a behavioral chamber. If the rats release within an allowed response window, they receive sugar water as reward. The data is recorded with electrodes spanning all cortical layers. All recorded neurons are from the Rostral Forelimb Area (RFA), which strongly contributes to planning and preparing for movements, with some having a direct connection to the Caudal Forelimb Area (CFA), responsible for motor execution.

Our contributions are threefold. First, we formalize NeuRL, a neural decoding method based on inverse action-value iteration. Second, we evaluate NeuRL in per-trial behavior prediction showing state-of-the-art performance. Third, we analyze the influence of neurons projecting from RFA to CFA by simulated inhibition within the NeuRL framework and real inhibition via viral manipulation. Our finding of similar response in real and simulated inhibition confirm that the intermediate representation of neural signals as immediate rewards offer a very promising direction for neural decoding methods.

\section{Background}

\subsection{(Inverse) Reinforcement Learning}
We model the task of neural decoding in the RL framework, where an agent (here a rat) acts in an environment as shown in \Cref{fig:ratrlsetting}.1. Following policy $\pi$ by applying action $a_t \sim \pi$ from $n$-dimensional action-space $\mathcal{A}$ in state $s_t$, it reaches some state $s_{t+1}\sim\mathcal{M}$ according to stochastic transition model $\mathcal{M}$ and receives scalar reward $r_t$ in each discrete time step $t$. The agent has to adjust its policy $\pi$ to maximize the expectation of long-term return $R(s_t) = \sum_{t'>=t} \gamma^{t'-t}r_{t'}$, where \mbox{$\gamma \in [0, 1]$} is a discount factor. The action-value function then represents the expected long-term value of an action when following policy $\pi$ thereupon, i.e. $Q^\pi(s_t,a_t)=\mathbf{E}_{a_{t'>t}\sim\pi,s_{t'>t}\sim\mathcal{M}}[R(s_t)|a_t]$. From the optimal action-value function $Q^*$ one can easily derive a corresponding optimal policy $\pi^*$ by maximization.

IRL recovers a reward function from observed trajectories from expert policy $\pi^\mathcal{R}$ under the assumption that the agent was (softly) maximizing the induced expected long-term return. Previous work solved this problem based on different approaches, such as 
Max Entropy IRL \citep{Ziebart2008MaximumEI}. %

\begin{figure}
    \centering
    \includegraphics[width=\textwidth]{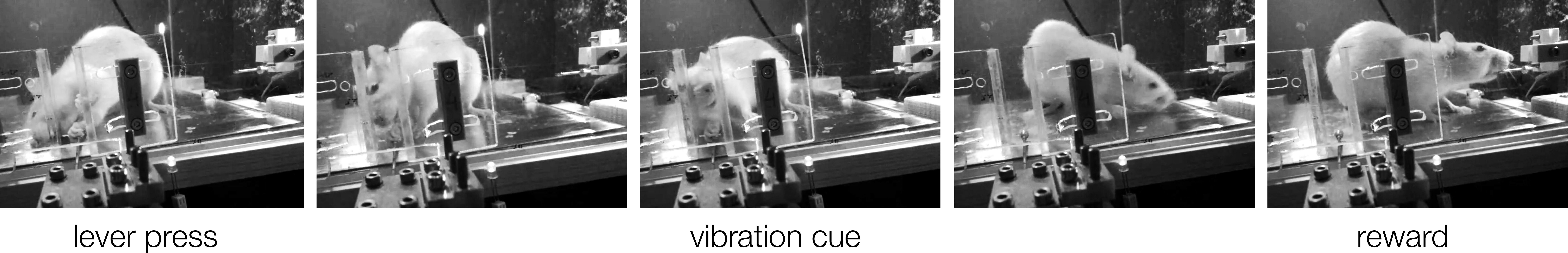}
    \caption{Successful trial of the response-preparation task in a behavioral chamber. The rat presses a lever until the vibration cue occurs, releases within \SI{0.6}{\second} and gets to the reward port.}
    \label{fig:ratsinchamber}
\end{figure}

\subsection{Action-value Iteration}

We focus on the case of finding the optimal policy via model-based Action-value Iteration. The Q-function, represented by a table with entries for every state and action, gets updated in every iteration $k$ based on the Bellman optimality equation with a given transition model $\mathcal{M}$: $$Q_k(s_t, a_t)\leftarrow r_t + \gamma\max_a\mathbf{E}_{s_{t+1}\sim\mathcal{M}}[Q_{k-1}(s_{t+1}, a))].$$

\section{Method}

In this section, we describe how to infer the scalar underlying reward function of a rat's behavior, the supervised approximation of this scalar reward as a weighted combination of neural signals and the neural decoding mechanism using the intrinsic reward function.

\subsection{Estimation of intrinsic reward}

We assume the rodent to softly maximize its measure of optimality which we define to be the expected cumulative sum of an unknown immediate reward function, i.e. the actions taken by the animal are samples from a Boltzmann distribution over its optimal action-values $Q^*(s,\cdot)$: \begin{align}\frac{e^{Q^*(s, a)}}{\sum_{A\in\mathcal{A}}e^{Q^*(s, A)}} \coloneqq \pi^\mathcal{R}(a|s),\end{align} for all actions $a\in\mathcal{A}$, and concomitantly:
\begin{align}
e^{Q^*(s, a)} &=\pi^\mathcal{R}(a|s)\sum_{A\in\mathcal{A}}e^{Q^*(s, A)} 
= \frac{\pi^\mathcal{R}(a|s)}{\pi^\mathcal{R}(b|s)}e^{Q^*(s, b)},
\label{eq:exp}
\end{align}
for all actions $b\in\mathcal{A}_{\bar a}$ where $\mathcal{A}_{\bar a}=\mathcal{A}\setminus\{a\}$. Following the derivations as proposed by \citet{NEURIPS2020_a4c42bfd}: 
\begin{align}
Q^*(s, a)=Q^*(s, b) + \log(\pi^\mathcal{R}(a|s)) - \log(\pi^\mathcal{R}(b|s)).
\label{eq:q}
\end{align}

Using the Bellman optimality equation in \Cref{eq:q}, the immediate reward of action $a$ in state $s$ can be expressed by the immediate reward of some other action $b\in\mathcal{A}_{\bar a}$, the respective log-probabilities and future action-values: 

\begin{equation}
    \begin{split}
        r(s, a) &= \log(\pi^\mathcal{R}(a|s)) - \gamma \max_{a'} \mathbf{E}_{s'\sim\mathcal{M}(s,a,s')}[Q^*(s', a')]\\
        &\qquad + r(s, b) - (\log(\pi^\mathcal{R}(b|s)) - \gamma \max_{b'} \mathbf{E}_{s''\sim\mathcal{M}(s,b,s'')}[Q^*(s'', b')]). 
    \end{split}
\end{equation}
Substituting the difference between the log-probability and the discounted action-value of the future state $s'$ as: \begin{align}\eta_s^a\coloneqq\log(\pi^\mathcal{R}(a|s)) - \gamma \max_{a'}\mathbf{E}_{s'\sim\mathcal{M}(s,a,s')}[Q^*(s', a')],\end{align} %
we can put the reward of action $a$ in state $s$ in relation to the reward of all other actions:
\begin{align}
r(s, a)  &=  \eta_s^a + \frac{1}{n-1} \sum_{b\in\mathcal{A}_{\bar a}}  r(s, b) - \eta_s^b.
\label{eq:final}
\end{align}

The resulting system of linear equations can be solved with least squares. We start by estimating the immediate reward for all terminal states and then go through the MDP in reverse topological order based on model $\mathcal{M}$. As can be seen in \Cref{subsec:imrew}, the Boltzmann distribution induced by the optimal action-value function on this learned reward is \emph{equivalent} to the arbitrary demonstrated behavior distribution (proof in \citep{NEURIPS2020_a4c42bfd}). IAVI thus returns a scalar \emph{intrinsic} reward function which precisely encodes the recorded behavior of subject rats as an intermediate result which can serve as supervised signal to learn a mapping from neural spiking. %

\subsection{Mapping of neural spiking to intrinsic reward}
\label{sec:mappingrtosig}
As second step, we map recorded neural spikes to the found intrinsic reward function in order to draw conclusions about the recorded behavior based on neural activity. We hence assume the immediate reward function to be a projection: \begin{equation}\hat{r}(\Phi(s),a) = \rho(\Phi(s)|\theta^\rho),\end{equation}
where $\rho$ is a parameterized function of features with parameters $\theta^\rho$, e.g. a linear combination or a neural network, and $\Phi(s)=(\Phi_1(s), \dots, \Phi_m(s))^\top$ the vector of $m$ features based on the recordings of $m$ neurons, such as the mean activity over all trials. We can fit parameters $\theta^\rho$ according to the class of function approximator, e.g. either by least squares or gradient descent, on the difference between reward $r(s,a)$ and prediction $\hat{r}(\Phi(s),a)$. The mapping can then be used to predict the resulting behavior based on neural spiking in new situations. %

\subsection{Neural Decoding From Intrinsic Reward}
The parameters $\theta^\rho$ of $\hat{r}(\Phi(s),a)$ are fitted to represent immediate reward $r(s,a)$ and hence the underlying behavior of the recorded rat as closely as possible. The found parameters can contribute to generalization to any \emph{arbitrary} neural spiking $\Psi(s)=(\Psi_1(s), \dots, \Psi_m(s))^\top$ which yields adjusted reward and action-values in each time step $t$:
\begin{equation}
    \begin{split}
        \hat r(\Psi(s_t),a_t) &= \rho(\hat \Psi(s_t)|\theta^\rho) \text{ and}\\
        \hat Q^*(\Psi(s_t),a_t) &= \max_\pi\mathbf{E}_\pi\left[\sum_{t'\geq t} \gamma^{t'-t} \hat r(\Psi(s_{t'}),a_{t'})\right].
    \end{split}
\end{equation}
From the optimal Q-function $Q^*(\Psi(s),a)$ based on features $\Psi(s)$, we infer the respective \emph{predicted} action-probabilities by: 
\begin{align}
    \hat \pi(a|\Psi(s))=\frac{e^{\hat{Q}^*(\Psi(s),a)}}{\sum_{A\in\mathcal{A}}e^{\hat{Q}^*(\Psi(s),A)}}.
\end{align}
In order to identify neurons or groups of neurons with particular relevance for a specific type of response, we can modulate their activity by modifying the respective features $\Phi_i(s)|_{1\leq i\leq m}$, keeping all other features fixed. Put differently, we can excite or inhibit certain neurons \emph{within the model} and make predictions about the response. The change in behavior between the ground truth based on the recorded spiking and the predicted response based on the modulated features provide insight over the possible individual impact of these neurons on cognitive processes. Furthermore, our model offers the possibility to learn the intrinsic reward of a rat along with the respective mapping from neural spiking to rewards based on \emph{recorded trials} in order to make predictions about behavioral response \emph{ad hoc} in active trials.

\section{Experiments}
In our experiments, we seek to find answers to the questions: (1) \emph{Is the immediate reward a good intermediate representation for neural decoding?} (2) \emph{How does NeuRL compare to the state of the art in per-trial action prediction?} (3) \emph{Are the responses predicted by NeuRL in line with real-world observations in inhibition experiments?} In the following section, we first describe the response-preparation task and the respective MDP formulation. Then, we compare the action probabilities as found in the data and predicted by the controller on basis of the learned reward function by IAVI. Lastly, we employ and compare NeuRL in per-trial behavior prediction and in the context of real-world inhibition.

\subsection{Response-Preparation Task}
\label{sec:task}
A total of six rats (two for the neural recordings used in our experiments and four for the real-world inhibition experiments) were placed into a behavioral chamber with one lever and a reward port (see \Cref{fig:ratsinchamber}). To complete the task and get the reward (sucrose water), the rats had to hold the lever for \SI{1.6}{\second} until a vibration to the paw occurs as a cue to release. The trial was considered correct if the rat released within \SI{0.6}{\second}. The rats were only rewarded for correct trials and were trained for 40 sessions over the course of two months. The subset of the data used for training our models comprises recordings of 30 neurons and 104 trials of rat 1 and 33 neurons and 184 trials of rat 2.\\%

\subsection{MDP Formulation}

\begin{figure}[h]
    \centering%
    \includegraphics[width=0.8\textwidth]{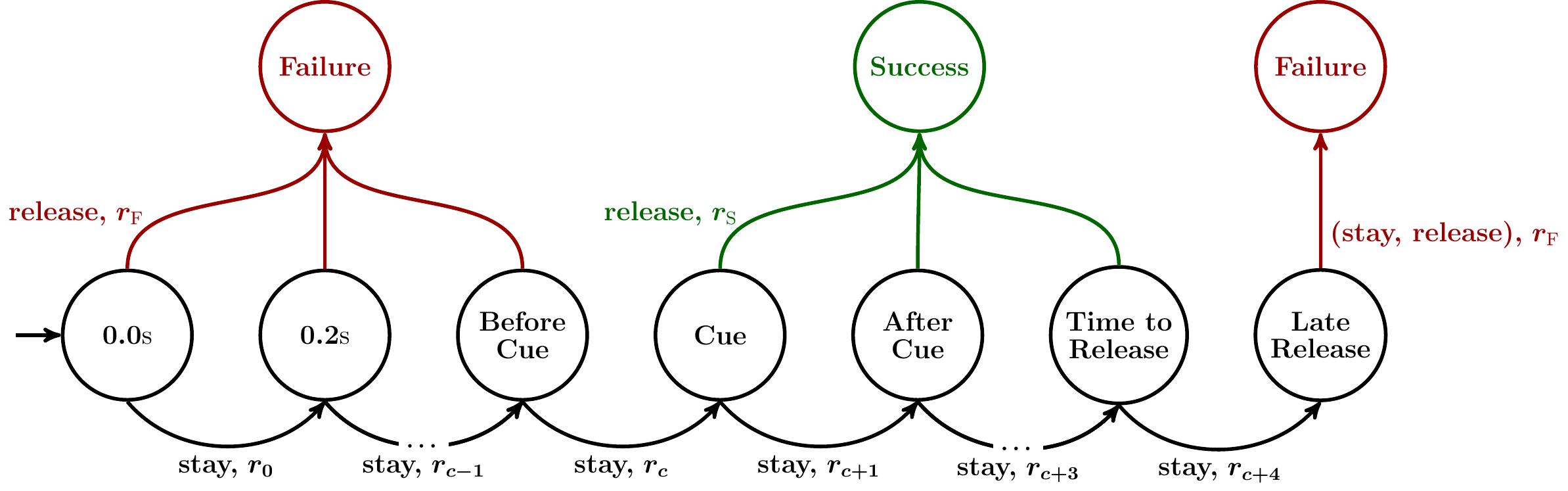}
    \caption{Transition graph for the MDP of the described response-preparation task. In the initial state, the rat presses the lever. If the rat does not release, it ends up in the next time step, where the time is discretized with \SI{0.2}{\second} steps. If the rat relases after the cue in a time span of \SI{0.6}{\second}, the trial was a success and it gets rewarded. $c$ denotes the running index over time steps before and after the cue (in our case $c=8$, representing \SI{1.6}{\second} with \SI{0.2}{\second} steps).}
    \label{fig:mdp}
\end{figure}

We model a simplified version of the response-preparation task as Markov Decision Process (MDP), where we consider the task after the press of the lever. The MDP is defined as a four-tuple $\langle\mathcal{S}, \mathcal{A}, \mathcal{M}, r\rangle$, where the set of states is defined by  $\mathcal{S}=\{\SI{0.0}{\second}, \SI{0.2}{\second}, \dots, \SI{1.2}{\second}\} \cup \{\text{Before Cue}, \text{Cue},\text{After Cue}, \text{After Cue}_1, \text{After Cue}_2, \dots,\text{Time to Release}, \text{Late Release}\}\cup \{\text{Suc-}$ $\text{cess},\text{Failure}\}$, discretizing the time into chunks of \SI{0.2}{\second}. In every state, the rat can pick an action from action space $\mathcal{A}=\{\text{stay}, \text{release}\}$. We define the MDP to have deterministic transitions. An overview is given in \Cref{fig:mdp}. In the following, we consider the reward function $r:\mathcal{S}\times\mathcal{A}\mapsto\mathbb{R}$ to be unknown.

\subsection{Reward Estimation via IAVI}
\label{subsec:imrew}
\begin{figure}[h!]
    \centering%
    \includegraphics[width=0.65\textwidth]{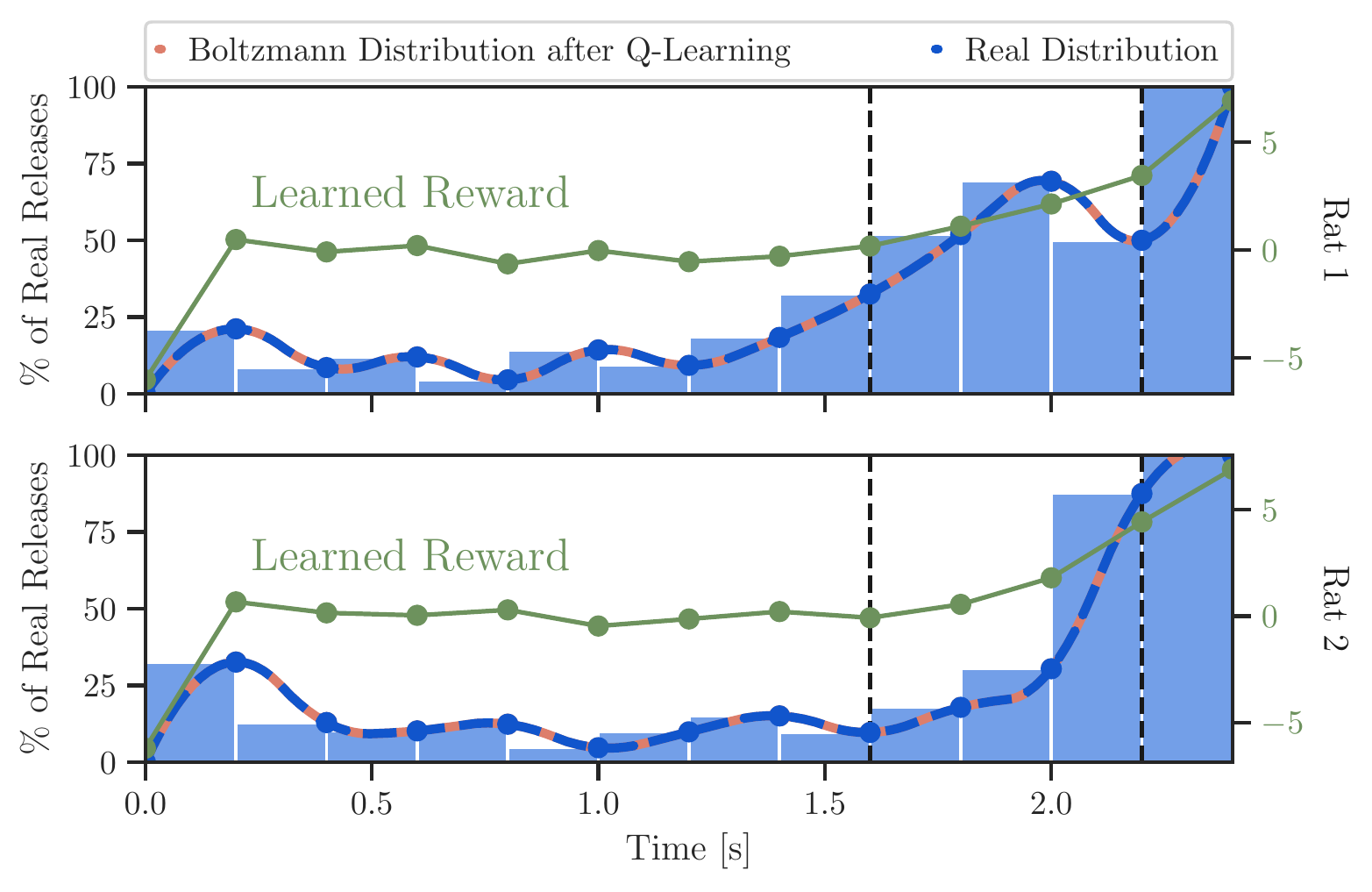}
    \caption{Probability of release for each time bin, learned reward and the resulting Boltzmann distributions after applying Q-learning on the reward for (top) rat 1 and (bottom) rat 2 over all trials. Dashed lines indicate the time span in which the rats ought to release.}
    \label{fig:realreleasesdist}
\end{figure}

To verify the correctness of the immediate reward found by IAVI, we first learn the intrinsic reward functions based on the recorded trajectories of rat 1 and rat 2 and the above defined MDP formulation. As can be seen in \Cref{fig:realreleasesdist}, the learned and real release distributions are identical, which shows that the scalar reward functions being found precisely explain the release distribution for each rat.

\subsection{Per-Trial Behavior Prediction from Neural Spiking}
\label{subsec:pertrialexp}

To study the performance for predicting actions of rats based on their neural signals in a trial with NeuRL, we use the neural spikings per time-step and trial as features and a neural network as function approximator for the reward. Since the resulting features are very sparse, we further append the time spent since trial initiation to the feature space. The learned intrinsic reward function is used to compute the corresponding release policy by applying action-value iteration on the reward. We compare NeuRL to  a random controller, logistic regression (LR) and non-linear classification via neural networks (NNC), which map directly from neural signal features to actions. Whenever a resulting controller assigns a probability of $>\epsilon$ (here we set $\epsilon=0.6$) to the action of release in a certain time step, we consider it a predicted release.  
We split the data set introduced in \Cref{sec:task} into different training and test sets using 10-fold cross-validation over all trials of a rat. For NeuRL and NNC, we optimized the hyperparameters with random search according to the configuration space in \Cref{tab:hpo} with 500 sampled configurations each.
\begin{table}[h!]
    \begin{subtable}{.5\linewidth}
      \centering
        \begin{tabular}{cc}
            \toprule
            Hyperparameter&Configuration Space\\
            \midrule
            \#updates&[5000, \textbf{\color{ForestGreen} 10000}, 20000]\\
            batch size&[16, 64, \textbf{\color{ForestGreen}256}]\\
            hidden dim&[\textbf{\color{ForestGreen}50}, 100, 200]\\
            num layers&[$\mathbf{\color{ForestGreen}2^*}, \mathbf{\color{ForestGreen}3^{**}}, 4$]\\
            learning rate&[$10^{-3}, \mathbf{\color{ForestGreen}10^{-4}}, 10^{-5}$]\\
            \bottomrule
        \end{tabular}
        \caption{Incumbents of {\color{ForestGreen}NeuRL}.}
    \end{subtable}%
    \begin{subtable}{.5\linewidth}
      \centering
        \begin{tabular}{cc}
            \toprule
            Hyperparameter&Configuration Space\\
            \midrule
            \#updates&[5000, \textbf{\color{Dandelion} 10000}, 20000]\\
            batch size&[\textbf{\color{Dandelion}16}, 64, 256]\\
            hidden dim&[$\mathbf{\color{Dandelion}50^{**}}$, 100, $\mathbf{\color{Dandelion}200^*}$]\\
            num layers&[$2, \mathbf{\color{Dandelion}3^{*}}, \mathbf{\color{Dandelion}4^{**}}$]\\
            learning rate&[$10^{-3}, \mathbf{\color{Dandelion}10^{-4**}}, \mathbf{\color{Dandelion}10^{-5*}}$]\\
            \bottomrule
        \end{tabular}
        \caption{Incumbents of {\color{Dandelion}NNC}.}
    \end{subtable} 
    \caption{Configuration space of hyperparameters. \textbf{Incumbent} for rat 1 ($^*$) and rat 2 ($^{**}$).}
    \label{tab:hpo}
\end{table}

\begin{figure}[h!]
    \centering
    \includegraphics[width=\textwidth]{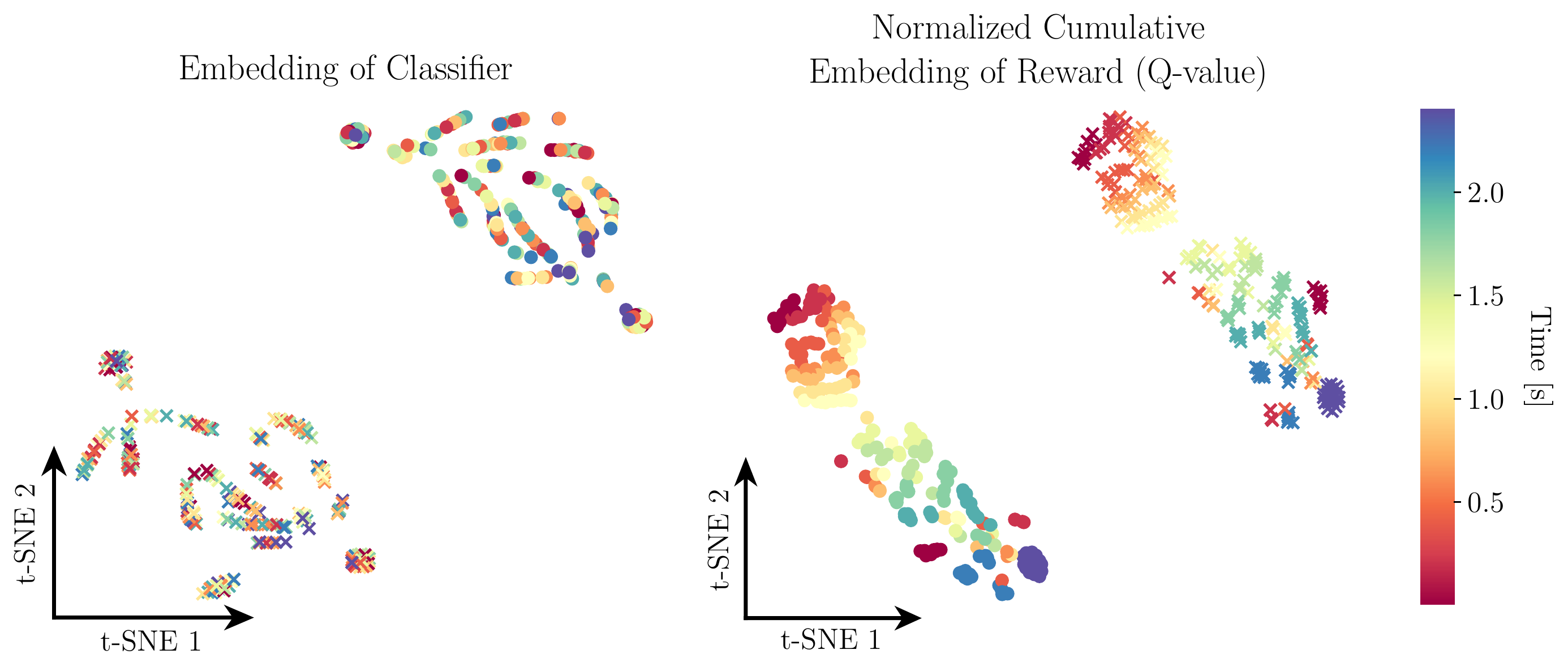}
    \caption{(left) Visualization of latent embeddings for the two actions \textit{stay} ($\circ$) and \textit{release} ($\times$) generated from the last hidden layer of the classifier (NNC). (right) Visualization of the normalized cumulative latent embeddings generated by the reward-model (as substitute for a Q-value). Our model preserves the temporal coherence of the task for the two actions \textit{stay} ($\circ$) and \textit{release} ($\times$) much better than the classifier on the left which is necessary for correct release prediction.}
    \label{fig:tsne}
\end{figure}

Results are shown in \Cref{tab:resultsclosedloop1}. NeuRL is able to correctly predict the releases in the test set by $36\%$ and $44\%$, respectively, for the two rats and exceeds the performance of all baselines by a large margin, also when considering near matches within one or two time steps. An intuition of why the immediate reward is a good intermediate representation can be gained from the visualization of the latent representation of the last hidden layers for the classifier (NNC) and NeuRL in \Cref{fig:tsne}. As substitute for a Q-value, we show the normalized cumulative embedding of the immediate reward. The latent representation of the neural features grounded in the learned immediate reward preserves the temporal coherence which stands in contrast to the latent embedding of the classifier. This brings light to the advantages of the proposed representation.

\aboverulesep=0ex
\belowrulesep=0ex
\renewcommand{\arraystretch}{1.0}
\begin{table}[t]
    \centering
    \begin{tabular}{c|ccc}
         \toprule
         &\multicolumn{3}{c|}{Rat 1}\\
         &Exact Match& Near 1 Match& Near 2 Match\\
         \midrule
         \textbf{\color{ForestGreen}NeuRL}&$\mathbf{0.36 (\pm0.11)}$ & $\mathbf{0.49 (\pm0.13)}$ & $\mathbf{0.59 (\pm0.09)}$\\
         \midrule
         {\color{Dandelion}NNC}& $0.21 (\pm0.09)$ & $0.28 (\pm0.12)$ & $0.37 (\pm0.17)$\\
         LR& $0.15 (\pm0.07)$ & $0.19 (\pm0.10)$ & $0.29 (\pm0.08)$\\
         Random&$0.04 (\pm0.07)$ & $0.20 (\pm0.13)$ & $0.29 (\pm0.15)$\\
         \bottomrule
    \end{tabular}\\
    \vspace{0.5cm}
    
    \begin{tabular}{c|ccc}
         \toprule
         &\multicolumn{3}{c}{Rat 2}\\
         &Exact Match& Near 1 Match& Near 2 Match\\
         \midrule
         \textbf{\color{ForestGreen}NeuRL}&$\mathbf{0.44 (\pm0.09)}$&$\mathbf{0.62 (\pm0.06)}$&$\mathbf{0.70 (\pm0.11)}$\\
         \midrule
         {\color{Dandelion}NNC}&$0.34 (\pm0.10)$&$0.46 (\pm0.09)$&$0.52 (\pm0.10)$\\
         LR& $0.33 (\pm0.09)$&$0.41 (\pm0.08)$&$0.47 (\pm0.10)$\\
         Random&$0.12 (\pm0.06)$&$0.38 (\pm0.07)$&$0.46 (\pm0.10)$\\
         \bottomrule
    \end{tabular}
    \caption{Mean prediction accuracy of release time step for 10-fold cross validation on rat 1 and 2.}
    \label{tab:resultsclosedloop1}
\end{table}

\aboverulesep=0ex
\belowrulesep=0ex
\renewcommand{\arraystretch}{1.0}
\begin{table}[h!]
    \centering
    \begin{tabular}{c|cc}
         \toprule
         &Rat 1&Rat 2\\
         \midrule
         \textbf{\color{ForestGreen}NeuRL}&$\mathbf{+23\%}$&$\mathbf{+26\%}$\\
         {\color{Dandelion}NNC}&$+16\%$&$+18\%$\\
         \bottomrule
    \end{tabular}
    \caption{Increase in mean prediction accuracy for rat 1 and 2 going from exact to near 2 matches.}
    \label{tab:resultsclosedloop3}
\end{table}

The larger increase in correct releases near two time steps, as shown in \Cref{tab:resultsclosedloop3}, further strengthens this point since most overlap in the latent embedding of NeuRL is between similar time steps (cf. \Cref{fig:tsne}).

\subsection{Simulation of Neural Inhibition and its Effect on the Rat's Behavior}

\begin{figure}[h]
    \centering
    \begin{subfigure}{0.45\textwidth}
    \begin{tikzpicture}[      
        every node/.style={anchor=south west,inner sep=0pt},
        x=1mm, y=1mm,
      ]   
     \node (fig1) at (0,0)
       {\includegraphics[width=\textwidth]{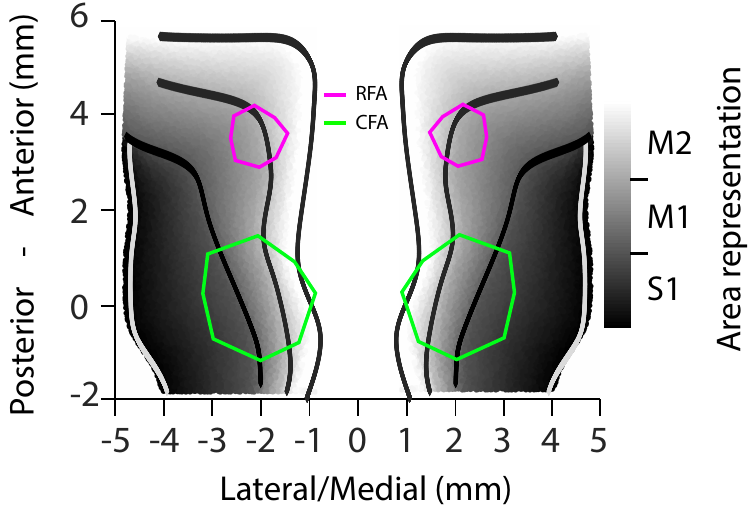}};
     \node (fig2) at (50,42)
       {\includegraphics[scale=0.33]{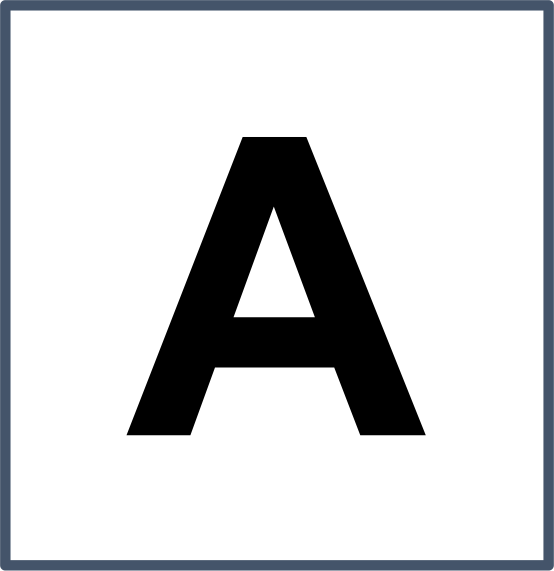}};  
    \end{tikzpicture}
    
    \captionlistentry{}
    \label{fig:brain}
    \end{subfigure}
    \hfill
    \begin{subfigure}{0.45\textwidth}
        \begin{tikzpicture}[      
        every node/.style={anchor=south west,inner sep=0pt},
        x=1mm, y=1mm,
      ]   
     \node (fig1) at (0,0)
       {\includegraphics[width=0.85\textwidth]{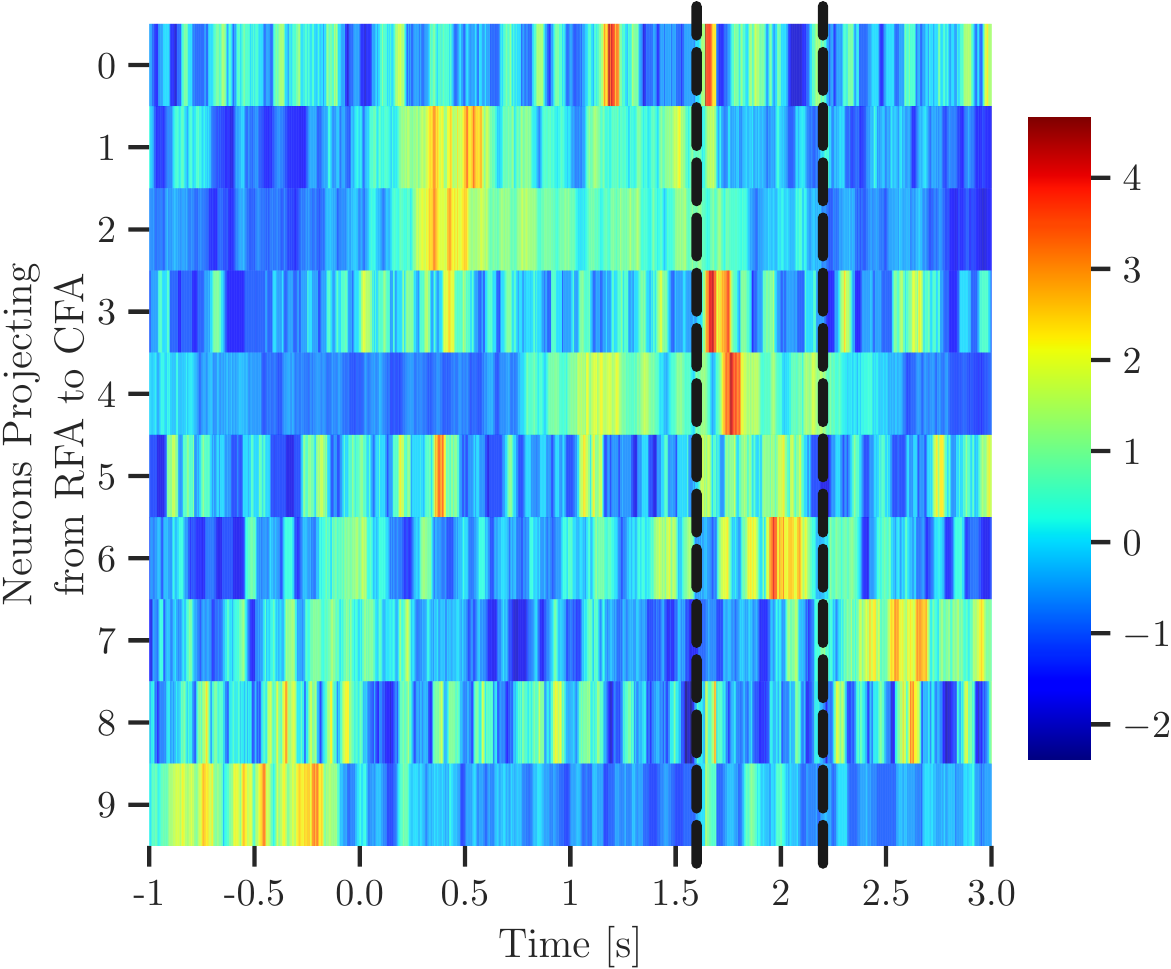}};
     \node (fig2) at (46,42)
       {\includegraphics[scale=0.33]{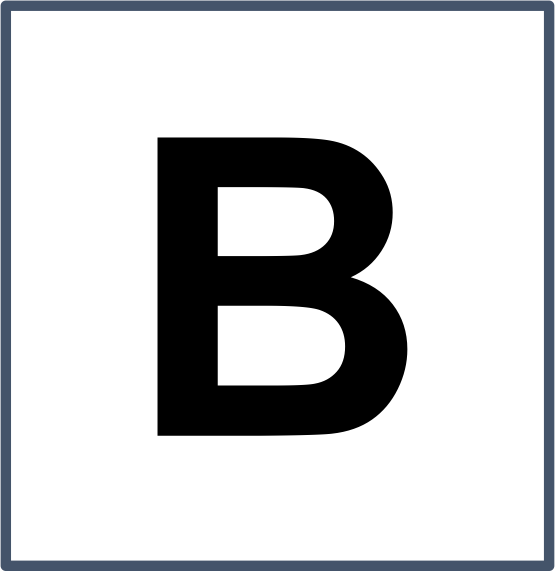}};  
    \end{tikzpicture}
    \captionlistentry{}
    \label{fig:heatmap}
    \end{subfigure}
    \caption{\fbox{\textbf{A}} Delineation of the Rostral (RFA) and Caudal (CFA) Forelimb Areas in a rat's brain according to \citet{NEAFSEY1982151} and \citet{doi:10.3109/08990229309028837}. \fbox{\textbf{B}} Z-score normalized firing rates of neurons projecting from RFA to CFA.}
    \label{fig:heatmapandpercentage}
\end{figure}

We study the influence of neurons projecting from RFA to CFA (cf. \Cref{fig:brain}) of which we identified 10 using optogenetic phototagging. The temporal pattern of the firing rate, as depicted in \Cref{fig:heatmap}, is surprisingly diverse even for a specific pathway. While the majority of the neurons (6/10, 60\%) increase the firing rate in the response period (indicated by dashed lines), about one third of the neurons (3/10, 30\%) have a higher firing rate in the hold period. %
Since most of the neurons are more active in the response window, we hypothesize that inhibition during this period has a significant effect on motor execution.

First, we simulate the effect of inhibition of these neurons within NeuRL and use the recordings of rats without viral manipulation. To simulate varying expected efficacy of viral manipulation, we sample subsets of the neurons projecting from RFA to CFA and set the respective features in $\Psi(s)$ within the allowed response window to zero (analogously to real-world inhibition experiments). We calculate the feature matrices accumulating the neural spikings by using an incremental mean over all trials for each rat (to aggregate all available information) and compute the weights $\theta^\rho$ via least squares, assuming a linear combination of the state features as described in \Cref{sec:mappingrtosig}. Then, as described above, we map the features to intrinsic rewards and compute the Q-values and corresponding stochastic Boltzmann policies. 

\begin{figure}[h]
    \centering
    \includegraphics[width=0.7\textwidth]{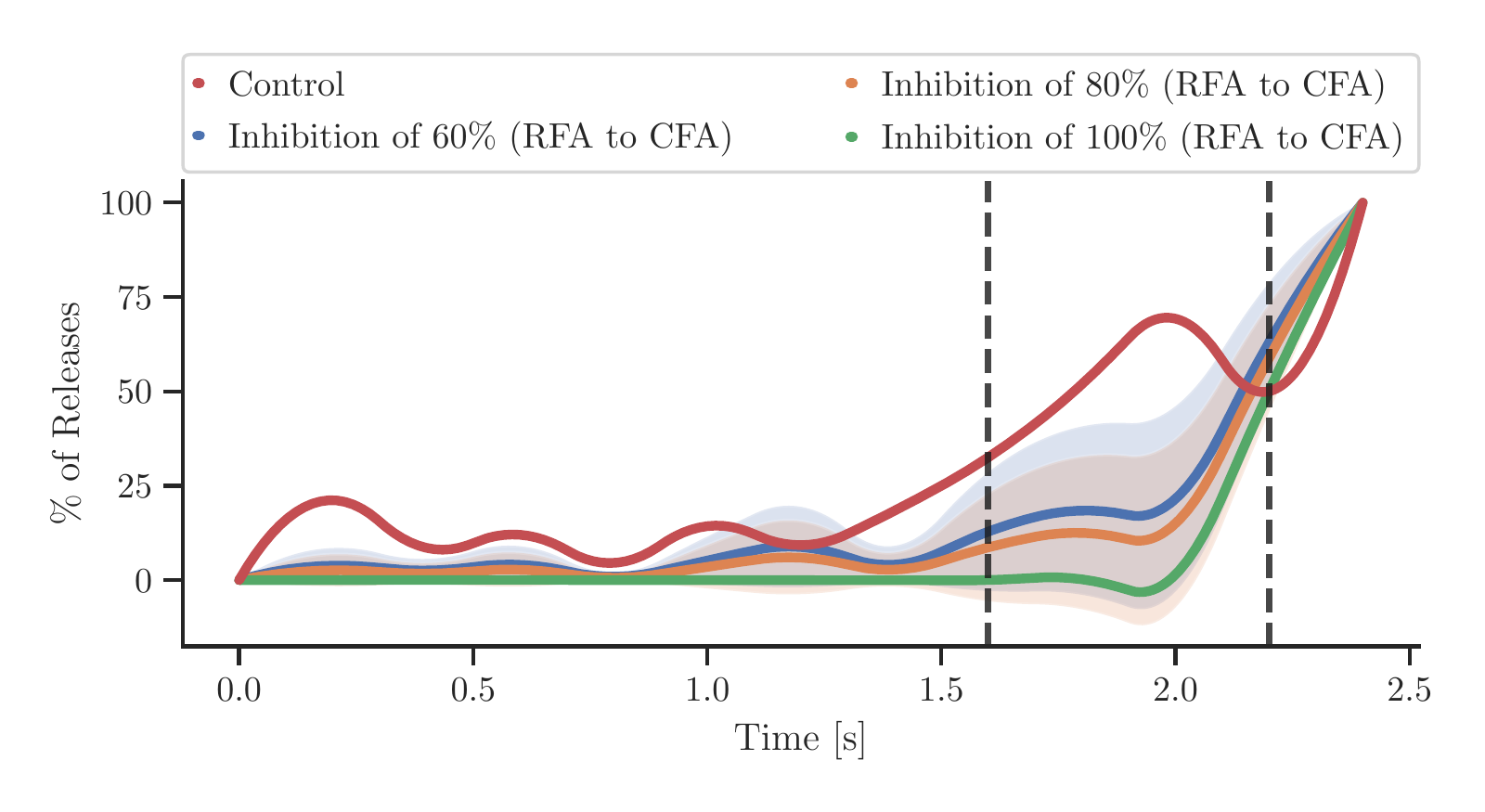}
    \caption{Release distribution of the rats (Control) and the resulting Boltzmann policies after applying Q-learning on the modified reward for different levels of simulated inhibition of neurons projecting from RFA to CFA. Dashed lines indicate the time span in which the rats ought to release.}
    \label{fig:rfatocfa}

\end{figure}

\Cref{fig:rfatocfa} shows the resulting release probabilities for different shares of inhibited neurons $(0.0, 0.6, 0.8$ and $1.0)$ in the response window between \SI{1.6}{\second} and \SI{2.2}{\second}. The inhibition causes late releases, as the probability of correct releases between \SI{ 1.6}{\second} and \SI{2.2}{\second} decreases with a higher proportion of inhibited neurons.
To evaluate our model, we consider trajectories of rodents solving the response-preparation task as defined in \Cref{sec:task} after neural inhibition via viral manipulation. To inhibit neurons \textit{in vivo}, we expressed the light gated inhibitory opsin enhanced Natronomonas pharaonis Halorhodopsin (eNpHr3.0 \citep{pmid18677566}) specifically targeting RFA to CFA projecting neurons in four trained rats. For this we injected a local Adeno Associated Virus (AAV)-based vector carrying the cre-dependent eNpHr construct into RFA and a retrograde traveling viral vector (retroAAV \citep{TERVO2016372}) providing cre recombinase into CFA. Thereby the opsin is only expressed in neurons projecting from RFA to CFA.  Experiments were conducted 12 weeks after injection to allow high levels of opsin expression. In 25\% of the trials, continuous light was delivered to RFA via optical fibers during the vibration cue.

\begin{figure}[h]
    \centering
    \begin{subfigure}{0.19\textwidth}
    \centering
    \raisebox{+0.56\height}{\includegraphics[width=\textwidth]{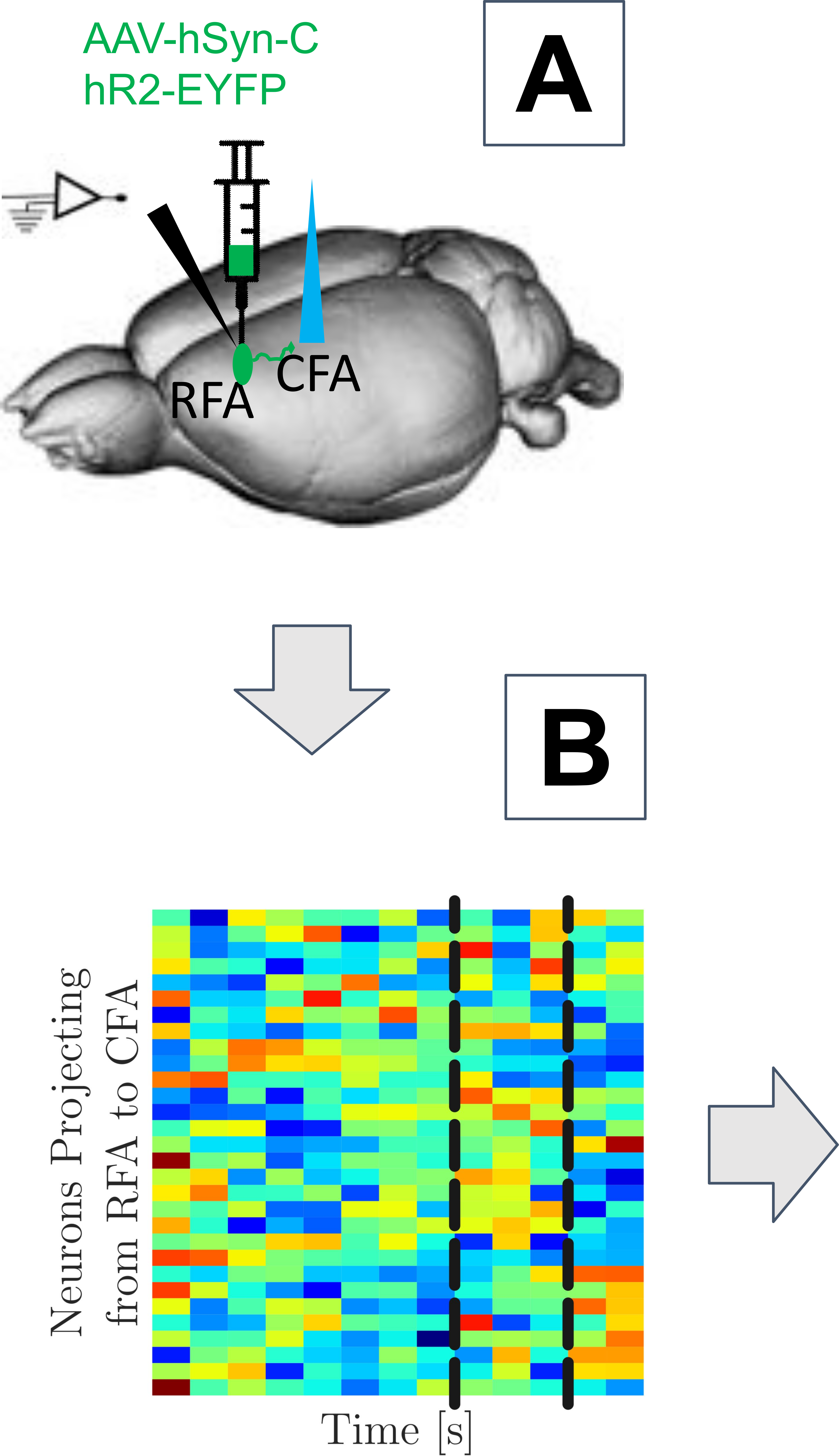}}
    \end{subfigure}
    \begin{subfigure}{0.6\textwidth}
    \centering
    \begin{tikzpicture}[      
        every node/.style={anchor=south west,inner sep=0pt},
        x=1mm, y=1mm,
      ]   
     \node (fig1) at (0,0)
       {\includegraphics[width=\textwidth]{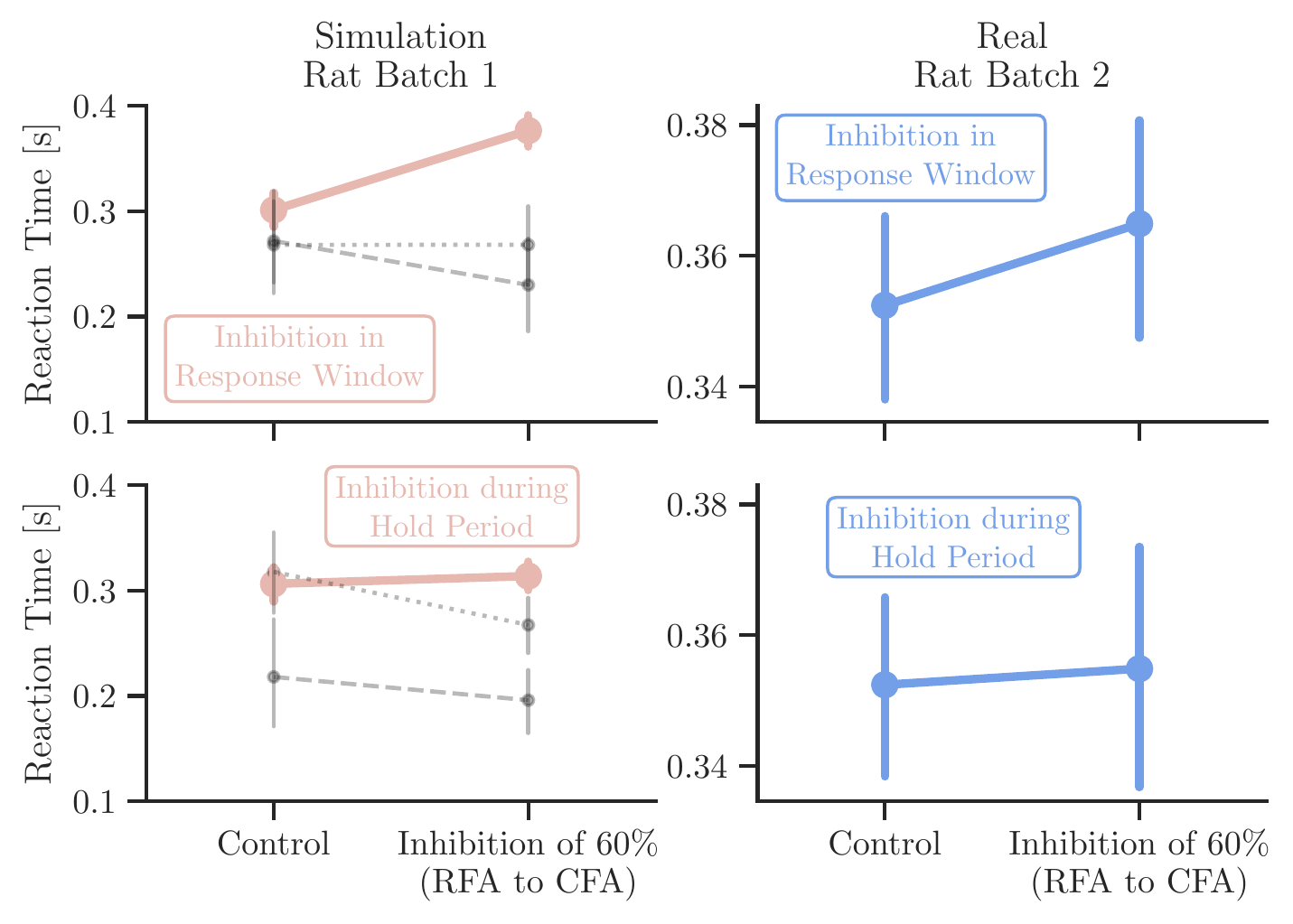}};
     \node (fig2) at (77,57)
       {\includegraphics[scale=0.33]{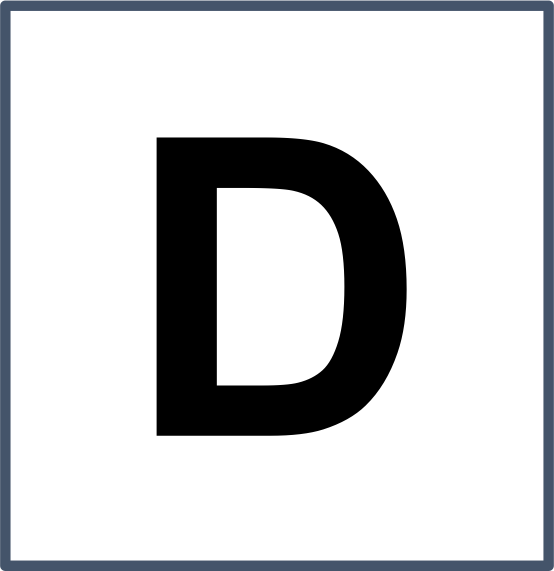}};  
    \end{tikzpicture}
    \end{subfigure}
    \begin{subfigure}{0.19\textwidth}
    \centering
    \raisebox{-1.85\height}{\includegraphics[width=\textwidth]{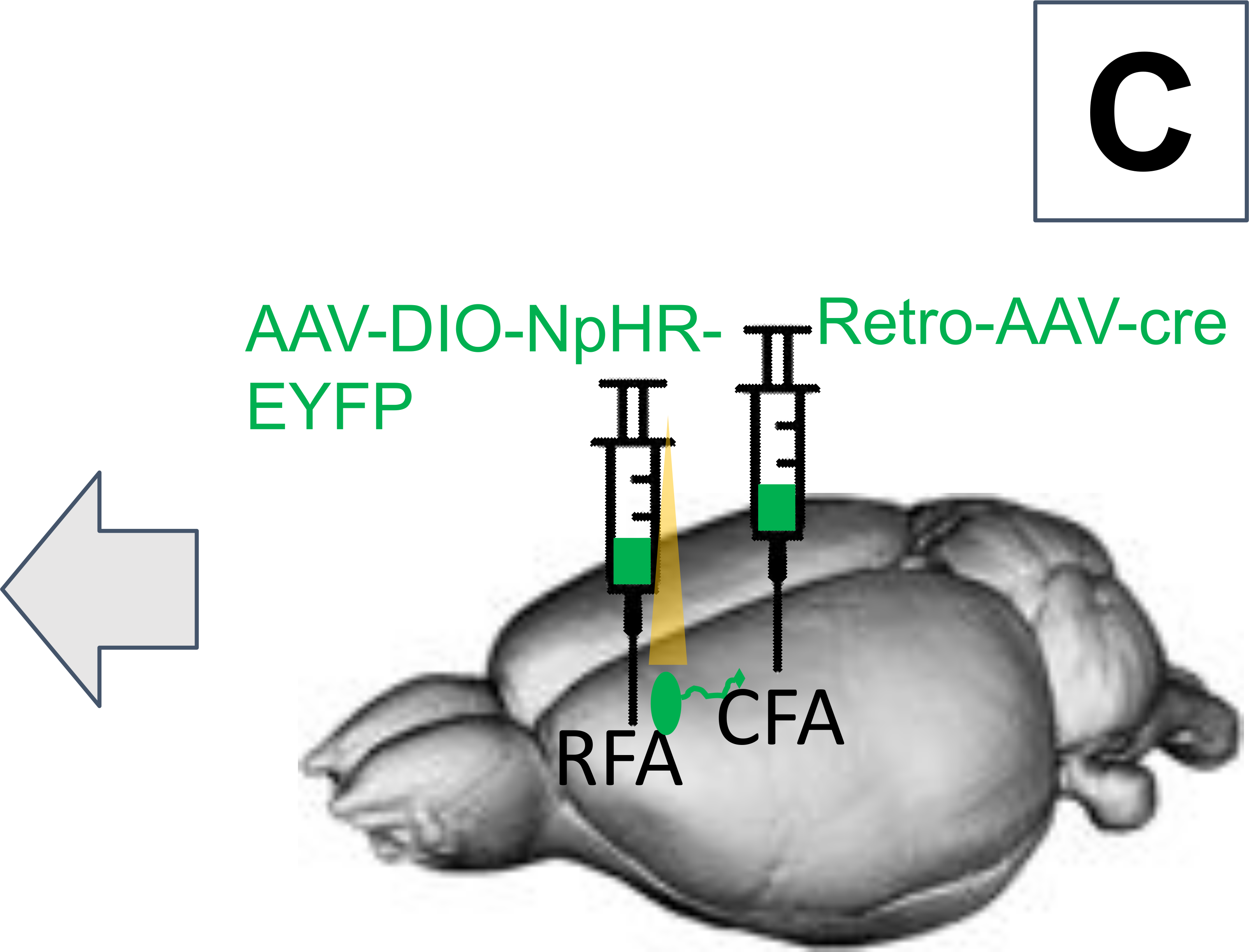}}
    \includegraphics[width=\textwidth]{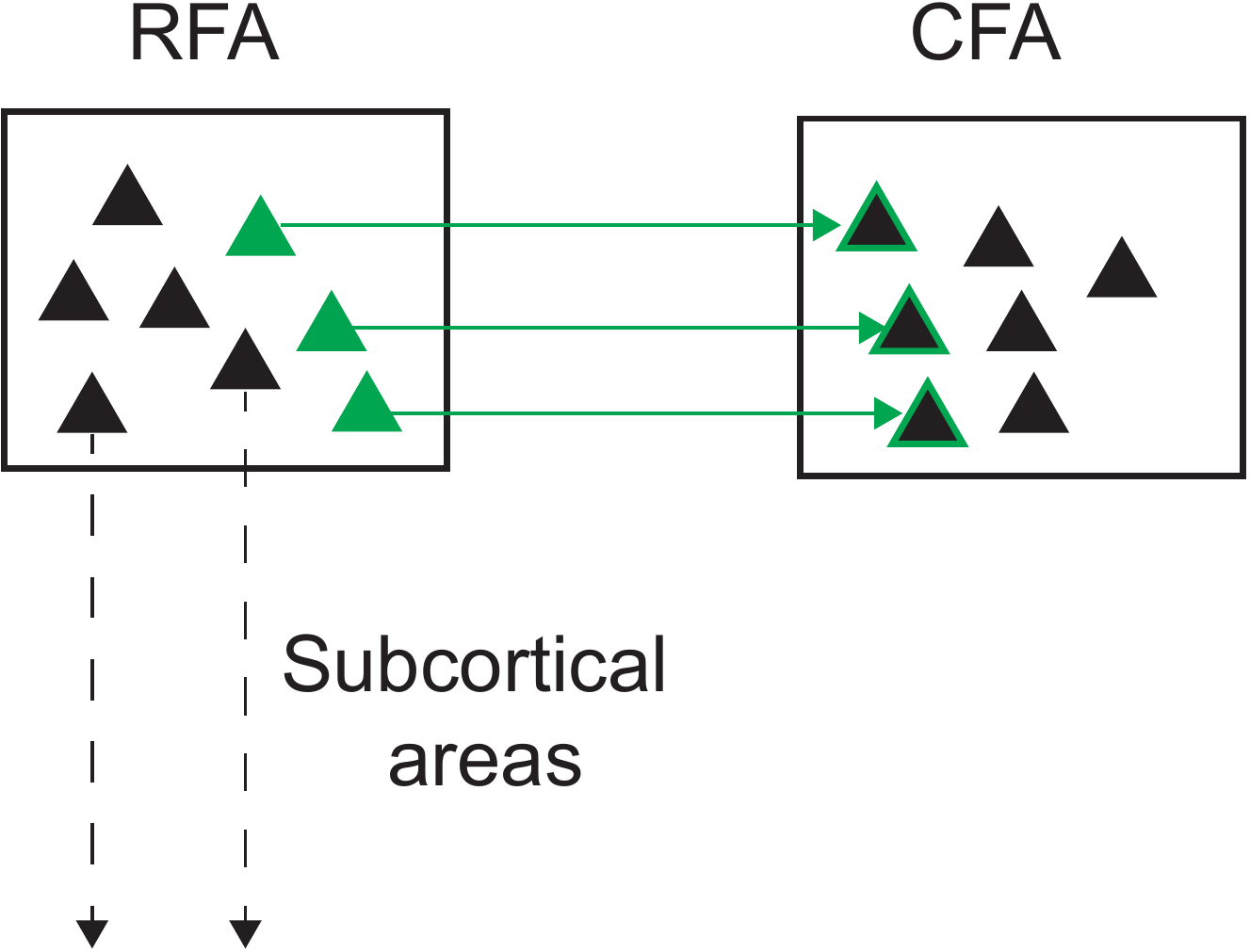}
    \end{subfigure}
    \caption{\fbox{\textbf{A}} Identification of relevant neurons via optogenetic phototagging. \fbox{\textbf{B}} Processing of neural recordings and extraction of neural spiking. \fbox{\textbf{C}} Viral manipulation of the pathway projecting from RFA to CFA. \fbox{\textbf{D}} Mean reaction times and standard error for (left) rat batch 1 without and with simulated inhibition of 60\% of RFA to CFA neurons and (right) rat batch 2 with and without real inhibition. (top) Within NeuRL (red) and in the real-world experiments (blue), the reaction time increases with inhibition in the response window. (bottom) There is no significant change in reaction time with inhibition during the hold period. Baselines are depicted in black for comparison, (\sampleline{dashed}) for linear regression and (\sampleline{dotted}) for NNC.}
    \label{fig:boxplot}
\end{figure}

In order to further put the performance of NeuRL in context of the current state of the art \citep{neuraldec}, we compare to logistic regression and NNC as described in \Cref{subsec:pertrialexp}. We train the baseline models on the recorded trials without viral manipulation on the basis of single time steps (analogously to our experiments in \Cref{subsec:pertrialexp}, the time spent since trial initiation is added to the feature space to account for the sparsity of per-time step features) and then use the predicted release probabilities according to the modified features $\Psi(s)$.

The resulting reaction times (time between cue and release in correct trials) for real and simulated inhibition with an efficacy of $60\%$ of the neurons projecting from RFA to CFA (corresponding to the efficacy of viral manipulation in practice) are summarized for all rats in \Cref{fig:boxplot}. The model provided by NeuRL captures both the tendency towards higher reaction times found in the real-world experiments of viral manipulation in the response window, as well as the absence of delay for inhibition during the hold period consistently for both rats. The difference in absolute numbers result from different subject rats for neural recording (basis for NeuRL) and real-world inhibition experiments. In contrast, logistic regression and non-linear classification are not able to reproduce the findings found in the in vivo experiments.

\section{Conclusion}
We introduced NeuRL, a three-step neural decoding method that first infers the true underlying immediate scalar reward function of a subject and then maps recorded neural spiking to this immediate reward in order to provide the possibility to decode unseen neural recordings thereafter. In simulated inhibition, our model was able to recover an effect of higher reaction times for the inhibition of neurons projecting from RFA to CFA shown in real-world experiments. In per-trial behavior prediction, our model achieved by far the best results, underlining the importance of reward prediction. Thus, our approach offers a novel and powerful interpretation tool for complex neuronal data, increasing the quality of behavioral predictions.

\bibliography{rat_rl}
\bibliographystyle{iclr2022_conference}

\end{document}